\documentclass[aps,prd,preprint,doublespaced,amsmath,amssymb]{revtex4}

\usepackage{graphicx}
\usepackage{dcolumn}
\usepackage{bm}

\begin{document}


\title{Gravity-Induced Geometric Phases and Entanglement in Spinors and Neutrinos: Gravitational Zeeman~Effect
}

\author{Banibrata Mukhopadhyay\email{bm@iisc.ac.in} and Soumya Kanti Ganguly
 \email{skganguly@physics.iisc.ernet.in}}
\affiliation{ Department of Physics, Indian Institute of Science
\\
Bangalore 560012, India\\
bm@iisc.ac.in, soumya09ganguly@gmail.com
}%


\begin{abstract}
	We show Zeeman-like splitting in the energy of spinors propagating in a background
gravitational field, analogous to the spinors in an electromagnetic field, otherwise termed the
Gravitational Zeeman Effect.~These spinors are also found to acquire a geometric phase, in~a
similar way as they do in the presence of magnetic fields. However, in a gravitational background,
the~Aharonov-Bohm type effect, in addition to Berry-like phase, arises. Based on this result, we~investigate
geometric phases acquired by neutrinos propagating in a strong gravitational field.~We~also explore
entanglement of neutrino states due to gravity, which could induce neutrino-antineutrino oscillation
in the first place. We show that entangled states also acquire geometric phases which are determined
by the relative strength between gravitational field and neutrino masses.
\\ \\


\end{abstract}
\maketitle

\section{Introduction}

It is well known that if the time dependence in the Hamiltonian 
arises through certain parameters, namely adiabatic parameters, then the system develops a non-dynamic phase,
called the Berry phase~\cite{berry}. Spinors propagating in the magnetic fields are known to acquire such a Berry phase.
Interestingly, a neutrino propagating through a medium also develops such 
a system, while the varying matter density corresponds to the adiabatic parameter. 
Importantly, although~originally Berry phase was found in the context of 
adiabatic, unitary and cyclic evolutions of time-dependent quantum systems,
later it was re-established for non-adiabatic, non-unitary and non-cyclic 
cases with its generalized definition~\cite{aa,sam,muksi}.

Several authors have
studied the geometric phases in neutrino oscillations. Although~it was argued in an earlier work 
that the Berry phase
plays no role in two-flavor neutrino oscillations in matter~\cite{naga}, the~work was restricted to a 
limited region in the parameter space. However, it was shown by exploiting the spin degree of freedom that 
the interaction of neutrinos with the transverse magnetic field can lead to a geometric effect~\cite{Vidal, Aneziris, Smirnov, Guzzo}.
Later on, it was argued~\cite{he} that the Berry phase can only appear in the presence of non-standard (e.g.,
R-parity violating supersymmetry) neutrino-matter interactions for the particular case of two-flavor
oscillations in matter. Essentially, all of the above papers argued that geometric phases do not arise in the two-flavor neutrino
oscillation probabilities with CP conservation in vacuum or in matter, in~the absence of any non-standard neutrino-matter
interactions. It was, however, furthermore argued~\cite{blasone} that even in the absence of CP violation, neutrinos
in two-flavor oscillation in vacuum in a period can acquire an overall phase consisting of a dynamical phase and
a phase depended on mixing angle only. The~second part of the phase, which is of geometric origin, 
was called Berry phase. Note that this phase does not arise due to slowly varying parameters leading to adiabatic
evolution, rather due to Schr\"odinger evolution of the system giving a closed loop in the Hilbert space.
As the phase is a global phase at the amplitude level, it does not appear in measurable quantities like 
probabilities of appearance or survival of neutrinos. These cyclic geometric phases were furthermore extended by
the later authors~\cite{wang} to obtain non-cyclic phases for two- and three-flavor neutrinos in vacuum, 
which remain unobservable because of the same reason as before. 
Also, the geometric phases for neutrinos propagating in varying magnetic fields have been 
reported~\cite{sudhir}.

It is interesting to note that~\cite{rajaram} the Berry phase has a connection to the phase discovered by 
Pancharatnam~\cite{pancha}. In~fact, both of the phases can be described under the same platform~\cite{sam}.
Unlike~the Berry phases obtained in the above work, it has however been established~\cite{mehta} that 
Pancharatnam phase can appear in detection probabilities and hence can be observed directly even in
an effective two-flavor approximation. {{There are many other explorations
over the years in various contexts of geometric phase and entanglement in
neutrinos, including those with CPT violation and fluctuating matter, 
non-linear refraction, magnetic field, dissipative matter, etc. } 
\cite{new1,new2,new3,new4,new5,new6,new7,new8}.} 

However, none of the works {above} considered the effects of gravity
in the calculations, {{except~one}~\cite{new8} {which considered Newtonian self-gravitational
interaction;}} whether the interaction of spinors and then neutrinos with gravitational field causes any
effect or not.~This issue particularly arises due to the fact that neutrinos interacting 
with background gravity may not preserve CPT~\cite{bm1,bm2}, which may be shown as a natural candidate for
governing the Berry phase, even in the evolution of neutrinos due to the split of dispersion energy between neutrino
and antineutrino. Indeed, within the pure standard model
of particle physics, the~neutrino oscillations cannot be understood and hence relaxing the CPT conservation
through gravitational interaction is one of the natural steps forward to beyond standard model. While
the Berry phase arises in the presence of non-standard matter-neutrino interactions, neutrino spin and    
magnetic field interactions, it is a natural question if the coupling between spin of neutrino and in general 
spinor and spin connection to the background gravity generates any geometric~effect.

Two-flavor neutrino oscillation in the background gravity has been discussed in various
astrophysical contexts. One of the current authors explored possible Lorentz and CPT 
violations in the neutrino sector in the presence of background gravity and its astrophysical
consequences~\cite{param,bmmpla,ujjal,bm1,bm2}. Earlier, the~analogy of solar neutrino oscillations
with the precession of electron spin in a time-dependent magnetic field was discussed~\cite{kim}.~Then based on the evolution of a statistical ensemble, oscillations for neutrinos from supernovae or
in the early universe in the presence of mixing and matter interactions in a thermal environment
were shown to be viewed in terms of precession~\cite{stodo}. It~was also observed~\cite{wudka} that spin flavor 
resonant transitions of neutrinos emanating from active galactic nuclei may occur in the vicinity of 
black holes due to gravitational effects and due to the presence of a large magnetic field. 
Interestingly, the~matter effects therein become negligible in comparison to gravitational~effects.

In the present paper, we start by recapitulating the origin of Berry phase in spinors in the presence of external 
magnetic fields in Section \ref{sec2}. Then we show the analogous effects in the presence of background gravitational fields, namely 
gravitational geometric phase in spinors in the same section. The subsequent plan is to apply this result in the neutrino sector. 
To do so, we first recapitulate the basic solutions of previous work discussing neutrino
oscillations in curved spacetime~\cite{bm1,bm2} in Section~\ref{solution}, which are used in subsequent sections. 
Based on these neutrino states evolving in the gravitational background,
we~explore any geometric (as well as dynamic) effect/phase arising due to gravity in Section \ref{sec4}. 
Subsequently, our aim is to explore the possible entanglement of neutrino states coupled with background gravitational 
field and to compute the geometric phase arising in their evolution in Section \ref{sec5}. Finally, we discuss how the 
geometric phases actually vary with gravitational field in Section~\ref{sec6} and summarize results in Section~\ref{sec7}. 

\section{\label{solution}Geometric Phases in the Presence of Electromagnetic and Gravitational~Fields}\label{sec2}
\vspace{-6pt}

\subsection{In Electromagnetic~Field}

{{The Dirac equation, describing dynamics of spinors, in~the presence 
of electromagnetic field 
and the underlying dispersion energy, Zeeman splitting and geometric/Berry phase are well-known.
However, for~the ease of understanding their similarities and also
dissimilarities with those in the gravitational field, which is the main
target here, we first recall the 
Dirac equation}} in the presence of electromagnetic field given by
\begin{eqnarray}
\left[i\gamma^\mu\left(\partial_\mu-ieA_\mu\right)-m\right]\psi=0,
\label{direm}
\end{eqnarray}
where the various components of $\gamma^\mu$, where $\mu=0,1,2,3$, 
are Dirac matrices with 
their usual meaning, $e$ is the electric charge, $m$ is the mass of the 
spinor and $A_\mu$ is the electromagnetic covariant 4-vector potential.
Here, we choose units $c=\hbar=1$.
For the non-trivial solution for $\psi$, the~energies/Hamiltonians of the spin-up and 
spin-down particles are 
given by
\begin{eqnarray}
(H+eA_0)^2=({\hat p}-e{\vec A})^2+m^2+e{\vec \sigma}\cdot{\vec B},
\label{direme}
\end{eqnarray}
where $A_0$ is the temporal component of $A_\mu$ which is basically the 
Coulomb potential, $\hat{p}$ is the quantum mechanical momentum operator $-i\nabla$, and~$\vec{\sigma}$ is the Pauli spin matrix. 
In the non-relativistic limit, where~$m^2$ is much larger than the rest of the terms in the R.H.S. of Equation~(\ref{direme}),
it reduces to
\begin{eqnarray}
H=-eA_0\pm\left[\frac{({\hat p}-e{\vec A})^2}{2m}+m+
\frac{e{\vec \sigma}\cdot{\vec B}}{2m}\right].
\label{direme2}
\end{eqnarray}

Apart from the split due to the positive and negative energy solutions, clearly 
there is an additional split in the respective energy levels. This is 
basically Zeeman-splitting governed by the term 
with Pauli's spin matrix, 
in the up and down spinors for the positive and negative energy spinors
induced
by magnetic fields, whether we choose relativistic or non-relativistic regimes.
The same governing term involved with $\vec{\sigma}$ is also responsible for the 
Berry phase if $\vec{B}$ is varying. 
For convenience, $\vec{B}$ is generally considered in the parameter space and is decomposed 
as $\vec{B}=|\vec{B}|\left(\hat{r}\sin\tilde{\theta}
\cos\tilde{\phi}+\hat{\tilde{\theta}}\sin\tilde{\theta}\sin\tilde{\phi}+\hat{\tilde{\phi}}
\cos\tilde{\theta}\right)$. Hence, the~Berry phase at a fixed $r$ is given by
\begin{eqnarray}
\Phi_g=i\oint\langle\psi|\nabla|\psi\rangle.d\vec{R}=\frac{\tilde{\phi}}{2}(1-\cos\tilde{\theta}),
\label{dirember}
\end{eqnarray}
where $\vec{R}\equiv(r,\tilde{\theta},\tilde{\phi})$, the~coordinate vector of the underlying 
Poincar\'e sphere. When $\vec{R}$ is constant, $\Phi_g=0$.

\begin{figure}
        \centering
        \fbox{\includegraphics[width=0.95\linewidth]{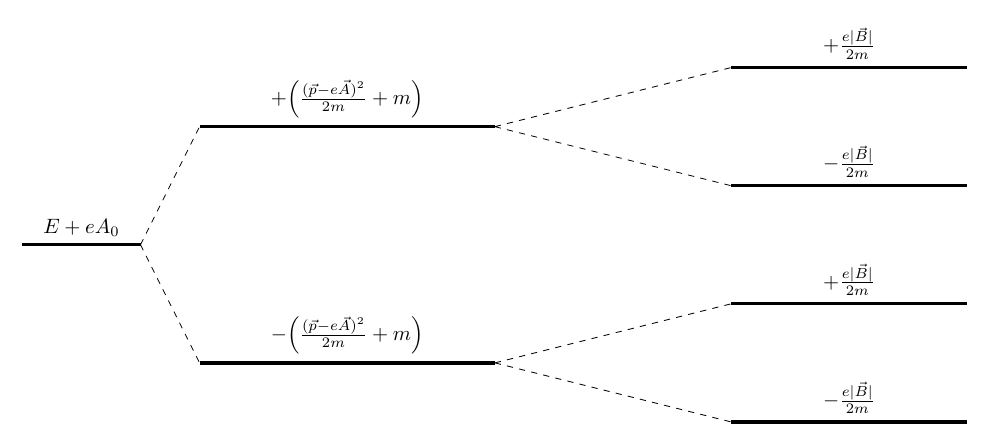}}
        \caption{
Zeeman-splitting in the electromagnetic case.
}
        \label{mage}
\end{figure}

Figure~\ref{mage} represents the energy splitting given by Equation~(\ref{direme2}). 
While the primary splitting corresponds to positive and negative energy solutions,
the secondary splitting corresponds to the interaction between the spin and magnetic~fields.

{{Recapitulation of all of the above results will be useful to explore and 
understand the consequences of the dynamics of spinors in the gravitational field.
As we will show below, although~there are certain similarities between the effects
of electromagnetic and gravitational fields to the spinors, there are some
unique consequences in the latter.}}

\subsection{In Gravitational~Field}

Dirac equation in the presence of background gravitational fields has already
been shown to have many consequences (see, e.g.,~the work by one of the
present authors~\cite{bm02,bmmpla,bm1,bm2,khushbooBM}) and is known to have the form 
(see, e.g.,~\cite{birrel,khushbooBM,kaku,bm2})
\begin{eqnarray}
\left[i\gamma^\mu\partial_\mu-m+\gamma^5\gamma^\mu B^g_\mu\right]\psi=0,
\label{dirg}
\end{eqnarray}
where $B^g_\mu$ is the gravitational covariant 4-vector potential (gravitational
coupling with the spinor), given by
\begin{eqnarray}
B^g_\mu=e^d_\mu B^g_d=\epsilon^{abc}_{~~d}~e_{b\lambda}\left(\partial_a e^\lambda_c
+\Gamma^\lambda_{\gamma\mu} e^\gamma_c e^\mu_a\right),
\end{eqnarray}
where $e^\lambda_c$-s and $\Gamma^\lambda_{\gamma\mu}$ are various components
of vierbeins and Christoffel connection with Greek and Latin indices 
respectively indicating curved and local flat coordinates, $\epsilon^{abcd}$
is the 4-dimensional Levi-Civita symbol
 and
$\gamma^5=\gamma_5=i\gamma^0\gamma^1\gamma^2\gamma^3$ as usual.
Here we do not repeat the calculation to obtain the reduced form 
of the Dirac equation given by Equation~(\ref{dirg}), which is available
in the existing literature, see, e.g.,~\cite{bmmpla,ujjal,schw} for details.
The form of Equation~(\ref{dirg}) is easy to understand in local inertial 
coordinates, where the Dirac $\gamma-$matrices and their relations are straight
forward. However, it is not difficult to explore Dirac equation in general 
curvilinear coordinates (see, e.g.,~\cite{obukhov}). Nevertheless, in~our
first exploration of the geometric phase of spinors and neutrinos in gravitational 
background, for~the convenience of developing the idea, we stick to the 
local inertial coordinates. This helps to compare results easily with
geometric/Berry phase arising in the presence of magnetic field without
losing any important physics. In~the future, we will report geometric/Berry phase
in gravitational field in global coordinates. Considering the problem in local
coordinates, in~brief, while~expanding various terms of the Dirac Lagrangian 
(and equation) in curved spacetime, one~obtains a Hermitian-like and anti-Hermitian-like parts, apart from the part already there in Minkowski 
spacetime. Hence, considering total Lagrangian consisting of that of particle
and anti-particle (and~corresponding equation),
the anti-Hermitian part drops out and one obtains Equation~(\ref{dirg}) 
given above (see~\cite{schw} for details). It can also be seen as the choice 
of an appropriate basis system~\cite{parker09}, particularly clearer when we explore it in
global non-inertial coordinates. Nevertheless, the~appearance 
of the anti-Hermitian-like part (which~need not always be anti-Hermitian, depending on the underlying spacetime)
is independent of the Hermitian-like term~\cite{bm02} that alone could lead to the 
axial-vector term given by Equation~(\ref{dirg}), which is the basic building block of the following 
discussion. Hence,~for~simplicity, here we do not consider the apparent
anti-Hermitian~term. 

Now, like the case of electromagnetic fields, for~the non-trivial solution of 
$\psi$ in the spacetime not explicitly dependent on time (except the case
where time dependence arises only via scale factor, like~in an expanding universe, 
and the exploration is in a particular epoch
or the interest is in the local-inertial coordinates), in~the local coordinates, 
the energies/Hamiltonians of the spin-up and spin-down particles 
from Equation~(\ref{dirg}) are 
given by
\begin{eqnarray}
(H-\vec{\sigma}\cdot\vec{B}^g)^2={\hat p}^2+{B_0^g}^2+m^2+(\hat{p}\cdot\vec{B}^g)+{\vec \sigma}\cdot\left[(\hat{p}B_0^g)+2B_0^g\hat{p}+(\nabla\times\vec{B}^g)-2\vec{B}^g\times\nabla\right],
\label{dirge}
\end{eqnarray}
where $B_0^g$ is the temporal component of $B_\mu^g$ and $\hat{p}=-i\nabla$.
In the regime of weak gravity and when $m^2$ is much larger than
the rest of the terms in the R.H.S. of Equation~(\ref{dirge}), it reduces to
\begin{eqnarray}
H=\vec{\sigma}\cdot\vec{B}^g\pm\left[\frac{{\hat p}^2+{B_0^g}^2+(\hat{p}\cdot\vec{B}^g)}{2m}+m
+\vec{\sigma}\cdot\left\{\frac{(\hat{p}B_0^g)+(\nabla\times\vec{B}^g)}{2m}+\frac{B_0^g\hat{p}-\vec{B}^g\times\nabla}{m}\right\}\right].
\label{dirge2}
\end{eqnarray}

Equation~(\ref{dirge2}) brings in a new effect due to the presence of an
axial-vector term in the Dirac equation in the gravitational background as 
opposed to the case with electromagnetic effects involved with a vector term
shown by Equation~(\ref{direme2}).
The new effect induced by the axial-vector involves the spin--momentum
coupling of the particle, apart from spin--gravity coupling, and~other
related terms, provided the background gravitational potential 
is non-zero. As~in general gravitational potential is not constant (even if 
locally gravitational field is constant), the eigenfunction of Equation~(\ref{dirge2})
cannot be plane-wave typed. However, if~$B^g_\mu$ is 
slowly varying and time-independent, then the solution can be of the form
\begin{equation} 
\psi=f(\vec{x},\vec{p})exp[i\vec{p}\cdot\vec{x}],
\label{solf}
\end{equation} 
where $f(\vec{x},\vec{p})$ is a slowly varying function. This leads the Hamiltonian to the form
\begin{eqnarray}
\nonumber
H=\vec{\sigma}\cdot\vec{B}^g\pm\left[\frac{{\vec p}^2-\frac{\nabla^2f(\vec{x},\vec{p})}{f(\vec{x},\vec{p})}
-\frac{2i(\nabla f(\vec{x},\vec{p}))\cdot\vec{p}}{f(\vec{x},\vec{p})}+{B_0^g}^2-i(\nabla\cdot\vec{B}^g)}{2m}+m\right.\\
\left.+\vec{\sigma}\cdot\left\{\frac{-i(\nabla B_0^g)+(\nabla\times\vec{B}^g)}{2m}+\frac{B_0^g\vec{p}-i\vec{B}^g\times\vec{p}
-iB_0^g\frac{\nabla f(\vec{x})}{f(\vec{x},\vec{p})}-\frac{\vec{B}^g\times\nabla f(\vec{x})}{f(\vec{x},\vec{p})}}{m}\right\}\right].
\label{dirge3}
\end{eqnarray}

As $f(\vec{x},\vec{p})$ is determined by the variation of background gravitational 
potential, we can suitably choose $f(\vec{x},\vec{p})$ in terms of $B_\mu^g$ and 
$\vec{p}$ in such a way that the above Hamiltonian is
Hermitian and terms involved with $f(\vec{x},\vec{p})$ are removed, given by
\begin{eqnarray}
H=\vec{\sigma}\cdot\vec{B}^g\pm\left[\frac{{\vec p}^2
+{B_0^g}^2}{2m}+m+\vec{\sigma}\cdot\left\{\frac{\nabla\times\vec{B}^g}{2m}
+\frac{B_0^g\vec{p}}{m}\right\}\right],
\label{dirge4}
\end{eqnarray}
where $f(\vec{x},\vec{p})$ satisfies
\begin{eqnarray}
\nonumber
\nabla^2f(\vec{x},\vec{p})+i\left(2\vec{p}+\frac{B_0^g}{m}\vec{\sigma}\right)\cdot\nabla f(\vec{x},\vec{p})+\vec{\sigma}\cdot\left(\vec{B}^g\times\nabla f(\vec{x},\vec{p})\right)\\
+i\left\{\nabla\cdot\vec{B}^g+\vec{\sigma}\cdot\left(\frac{\nabla B_0^g}{2m}+
\frac{\vec{B}^g\times\vec{p}}{m}\right)\right\}f(\vec{x},\vec{p})=0.
\label{feq}
\end{eqnarray}

Note that the solution for $f(\vec{x},\vec{p})$ will turn out to be complex
and its imaginary part may need to be adjusted by modifying $\vec{p}$ of the solution.
Obviously, this is one of the possible solutions and is a gauge choice, not 
unique. However, this will suffice for the present purpose when the aim is to
show the existence of geometric phase and a possible new effect in gravitational~background.

From Equation~(\ref{dirge4}), there is a two-fold split in dispersion energy, governed by two terms associated with the Pauli 
spin matrix, between~up and down spinors for the positive and negative energy 
spinors induced
by gravitational fields, whether the field is weak or strong.
The same governing terms are also responsible for the Berry phase, as~is for electromagnetic 
fields, which in the parametric space with coordinates $\vec{R}\equiv(r,\tilde{\theta},\tilde{\phi})$ at a fixed $r$ is given by
\begin{eqnarray}
\Phi_g=i\oint\langle\psi|\nabla|\psi\rangle\cdot d\vec{R}=\frac{\tilde{\phi}}{2}(1-\cos\tilde{\theta}),
\label{dirember}
\end{eqnarray}
which was briefly introduced by one of us earlier~\cite{bookmy}.

The first term in the curly bracket of Equation~(\ref{dirge4}) 
is the magnetic-equivalent
contribution from gravitational field. Interestingly, even if $B^g_\mu$ is constant but non-zero, 
but $\vec{p}$ is varying---at least changing direction due to whatever
reason, e.g.,~the presence of constant magnetic field which however does not 
produce any geometric phase---$\Phi_g$ survives, as~seen from the last term in Equation~(\ref{dirge4}). 
The~contribution from the first term in Equation~(\ref{dirge4}) adds up to $\Phi_g$
if $B^g_\mu$ varies (i.e., potential varies but field need not necessarily vary) 
and that from the second last term if $\nabla\times \vec{B}^g$
varies. Hence,~while~in electromagnetic fields the magnetic potential and hence field has
to be varying, in~gravitational field even the constant (but non-zero) 
gravitational potential (and also constant gravitational field) still would
produce geometric/Berry~phase.

\begin{figure}
        \centering
        \fbox{\includegraphics[width=0.95\linewidth]{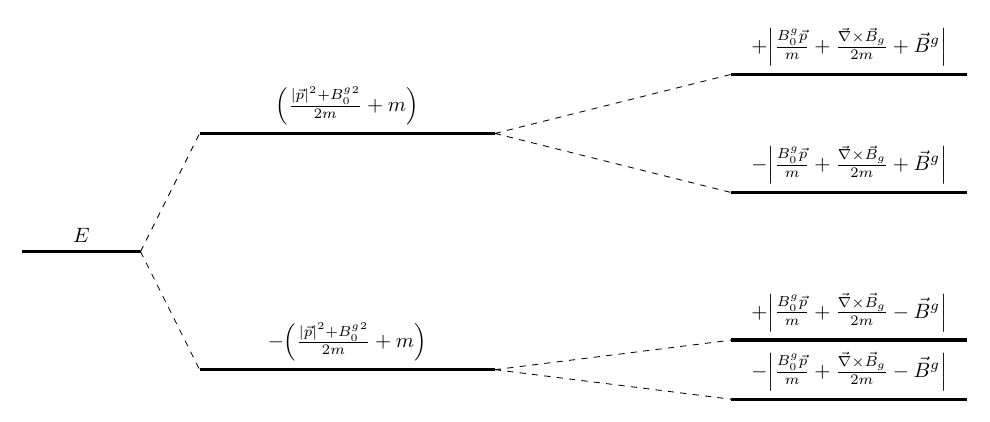}}
        \caption{Gravitational ``Zeeman-splitting''. 
}
        \label{grave}
\end{figure}

Figure~\ref{grave} represents the energy splitting given by Equation~(\ref{dirge4}),
which was already introduced briefly by one of us~\cite{bookmy,khushbooBM} {{(note,
however, a~typo in the notation of the first splitting in those works).}}
Here, the splittings are different to those in electromagnetic case. 
The total splittings are involved with the interaction between the spin, 
background gravitational potential and gravitational field. 
Hence, the~gravitational
``Zeeman-effect'' appears to be different to the conventional electromagnetic 
Zeeman-effect. While the electromagnetic Zeeman-effect and geometric phase 
depend on the underlying magnetic field only, their gravitational counter-parts
in general involve both gravitational potential and field. Therefore, gravitational
effects reveal an Aharonov-Bohm type effect in addition to Berry-like phase.
Nevertheless, the~total energy of the system
of particles remains conserved in both electromagnetic and gravitational cases 
(which indeed should be in the time-independent spacetime). Also,
in the local inertial frame, at~a given epoch if the process is
considered in expanding universe, gravitational potential $B^g_\mu$ 
appears to be constant, acting as a background~effect.

Note that $B^g_\mu$ can be computed for various spacetime metrics,
as given by previous work~\cite{bm1,bm2,param,bmmpla,ujjal,bm02}.
In order to have a non-zero $B^g_\mu$, spherical symmetry has to be broken
and hence in the Schwarzschild geometry (and hence for the spacetime 
around a non-rotating black hole), it vanishes. On~the other hand,
in the Kerr geometry (and hence for the spacetime around a rotating black hole),
it survives independent of the choice of coordinates: in the Boyer-Lindquist 
as well as the Kerr-Schild~\cite{param,bm1}. Also, it survives in other natural
spacetimes breaking spherical symmetry, e.g.,~in early universe 
under gravity wave perturbation, the~Bianchi II, VIII and IX anisotropic 
universe, in~the Fermi-normal coordinates up to second order correction~\cite{bm1,bm2,ujjal,bm02}. In~the Kerr-Schild coordinates, the~temporal
part of gravitational potential reads as~\cite{bm1,bm2}
\begin{equation}
B^g_0=-\frac{4a\sqrt{M} z}{\bar{\rho}^2\sqrt{2r^3}},
\end{equation}
where $\bar{\rho}^2=2r^2+a^2-x^2-y^2-z^2$, $r$ is the radial coordinate of
the system and $M$ and $a$ are respectively the mass and the
angular momentum per unit mass of the black hole. Naturally, $B^g_0$
survives (and~is varying with space coordinates) for any non-spinning black hole leading to gravitational
Zeeman effect and Berry phase independent of spatial part $\vec{B}^g$. Similarly, non-zero
$\vec{B}^g$ leads to gravitational Zeeman effect and Berry phase (when 
$\vec{B}^g$ has to be varying as well), independent of 
$B^g_0$. In~the Bianchi II spacetime with, e.g.,~even equal scale-factors
in all directions, $B^g_0$ survives as~\cite{ujjal,bm1}
\begin{equation}
B^g_0=\frac{4+3y^2-2y}{8+2y^2},
\end{equation}
leading to both gravitational Zeeman splitting and Berry phase 
even though $\vec{B}^g=0$. 

Also see similar exploration in varied
contexts~\cite{kost,nick1,nick2,harm}. Hence, the axial vector term in Equation~(\ref{dirg}) and the related terms involving with ${\vec \sigma}$ in Equation~(\ref{dirge2})
contribute as long as the spacetime naturally has some handedness, independent
of the choice of~coordinates.

\section{\label{solution}Neutrino States in the Presence of Gravitational~Field}
{{As discussed in the Introduction, over~the years there have been several
explorations of neutrinos in 
the presence of gravitational field. For~the present purpose, we particularly
use some of earlier results}~\cite{bm2}{ obtained by one of the present authors
and further modify as required.}}

\subsection{Neutrino--Antineutrino~Mixing}

Recalling the work by Sinha and Mukhopadhyay~\cite{bm2} describing 
the mixing of neutrinos ($\psi$) and antineutrinos ($\psi^c$) in the presence of 
gravitational coupling, based on the formalism discussed above,
let us write down the mass eigenstates $\nu_1$ and $\nu_2$ for a particular flavor at $t=0$
\begin{subequations}\label{mix}\begin{eqnarray}
|\nu_1(0)\rangle~=~\cos \theta~|\psi^c(0)\rangle~+~e^{i\phi}~\sin
\theta~|\psi(0)\rangle \\
|\nu_2(0)\rangle~=~-\sin \theta~|\psi^c(0)\rangle~+~e^{i\phi}~\cos \theta~|\psi(0)\rangle, \end{eqnarray}
\end{subequations} 
when \begin{equation} \tan\theta~=~\frac{m}{B_0+\sqrt{B_0^2+m^2}} \label{eqn20},\,\,\,\,\,\phi=arg(-m),
\end{equation} 
where $B_0$ is the gravitational scalar coupling potential and $m$ the Majorana mass of the neutrino. Henceforth, by~$B_\mu$ we will mean $B_\mu^g$ itself, 
defined in the previous section, 
in order keep the same notation as of previous papers.
The large $B_0$ corresponds to $\theta\rightarrow 0$, hence no mixing and 
thence no oscillation.
However, at~an arbitrary time $t$ the mass eigenstates are
\begin{subequations}\label{mixt}\begin{eqnarray}
|\nu_1(t)\rangle~=~\cos \theta~e^{-iE_{\psi^c} t}~|\psi^c(0)\rangle~+~e^{i\phi}~\sin
\theta~e^{-iE_\psi t}~|\psi(0)\rangle \\
|\nu_2(t)\rangle~=~-\sin \theta~e^{-iE_{\psi^c} t}~|\psi^c(0)\rangle~
+~e^{i\phi}~\cos \theta~e^{-iE_\psi t}~|\psi(0)\rangle, \end{eqnarray}
\end{subequations} 
where $E_\psi$ and $E_{\psi^c}$ are dispersion energies of neutrino and anti-neutrino respectively, in~local coordinates neglecting their variation, 
given by (see also, e.g.,~\cite{khushbooBM})
\begin{eqnarray}
\nonumber
E_\psi=\sqrt{(\vec{p}-\vec{B})^2+m^2}+B_0,\\
E_{\psi^c}=\sqrt{(\vec{p}+\vec{B})^2+m^2}-B_0,
\label{energy}
\end{eqnarray}
where $\vec{B}$ is the gravitational vector coupling potential and $\vec{p}$
the momentum of the neutrinos.
In the absence of a gravitational field, neutrinos and antineutrinos mix
in the same angle, hence there is no neutrino--antineutrino oscillation.
This is indeed in accordance with the experimental finding~\cite{exp}.

The underlying oscillation length can also be recalled, for~ultra-relativistic and non-relativistic
(or~weakly gravitating) neutrinos, as~\begin{equation} \lambda=\frac{\pi}{B_0-|\vec B|}\,\,\,{\rm and}\,\,\,
\frac{\pi}{B_0}\,\,\,{\rm respectively}.
\end{equation} 

This depends only
on the strength of the gravitational field. If~we consider
neutrinos to be coming out off the inner accretion disk, of~few factors times Schwarzschild radii, around a
spinning black hole of mass $M=10M_\odot$, where $M_\odot$ is the mass of Sun, then
$B_0-|\vec{B}|=\tilde{B}=10^{-19}$ GeV~\cite{bm1}, which clearly argues
neutrinos not to be influenced by the gravitational field of black hole 
as \mbox{$m>> 10^{-10}$ eV.} However,
for $\tilde{B}<<m$ the energy difference also turns out to be $2B_0$ (which is $2B_0-2|\vec{B}|$
for ultra-relativistic neutrinos), which leads
to $\lambda\sim 10$ km. If~the disk is around a supermassive black
hole of $M=10^8M_\odot$, i.e.,~in an Active Galactic Nucleus (AGN), then $\lambda$ may increase
to $10^{8}$~km. 
Therefore, an~oscillation between mass eigenstates may complete from a few 
factors to hundreds of Schwarzschild radii in the disk, 
depending upon the size of the inner edge where neutrinos come out and 
angular momentum of the black hole. This may produce copious
antineutrinos over neutrinos and may cause overabundance of neutrons
and positrons, {{which may further have consequences in core-collapse
supernovae.} }
However, neutrinos around a primordial black hole
of mass $M_p\sim 10^{17}$ gm \footnote{Note that the corresponding temperature
$T_p\sim 10^{-20}\,M_\odot/M_p$ GeV~\cite{st}.} in the site of temperature 
$1$ MeV are ultra-relativistic and could lead to an
oscillation length as small as \mbox{$\lambda\sim 10^{-16}$ km}. 
Note that $(\nu_1, \nu_2)$ is just the transformed spinor of original $(\psi^c,\psi)$.

\subsection{Mixing of Mass~Eigenstates}

Following previous work~\cite{bm2}, neutrino and anti-neutrino states can also be
in principle written as a linear combination of suitable mass states at $t=0$ as
\begin{subequations}\label{eqn21}\begin{eqnarray}
|\psi^c(0)\rangle~=~\cos \theta~|\nu_1(0)\rangle~-~~e^{i\phi}~\sin \theta~|\nu_2(0)\rangle \\
|\psi(0)\rangle~=~\sin \theta~|\nu_1(0)\rangle~+~~e^{i\phi}~\cos \theta~|\nu_2(0)\rangle
\end{eqnarray} \end{subequations}
and at an arbitrary time $t$ as
\begin{subequations}\label{eqn22}\begin{eqnarray}
|\psi^c(t)\rangle~=~\cos \theta~e^{-iE_1 t}~|\nu_1(0)\rangle~-~~e^{i\phi}~\sin \theta~e^{-iE_2 t}
~|\nu_2(0)\rangle \\
|\psi(t)\rangle~=~\sin \theta~e^{-iE_1 t}~|\nu_1(0)\rangle~+~~e^{i\phi}~\cos \theta~e^{-iE_2 t}~|\nu_2(0)\rangle,
\end{eqnarray} \end{subequations}
where in ultra-relativistic limits the energies of the mass eigenstates are
\begin{eqnarray}
E_{(1,2)}\approx|\vec{p}|+|\vec{B}|+\frac{m_{(1,2)}^2}{2|\vec{p}|},
\end{eqnarray}
when the corresponding masses
\begin{eqnarray}
m_1=m_n-\sqrt{B_0^2+m^2},\,\,\,\,m_2=m_n+\sqrt{B_0^2+m^2},
\label{mas}
\end{eqnarray}
in the presence of lepton number conserving mass ($m_n$) and 
violating mass ($m$). Here, $|\vec{p}|\sim E$, the~mean energy of the neutrinos, and~$\theta$ and $\phi$ are
the same as in Equation~(\ref{eqn20}). 
The oscillation length is
\begin{equation} 
\lambda_n = \frac{\pi E}{m_{n}\sqrt{B_{0}^2+m^2}+ \vert \vec{B}\vert E}, 
\end{equation}
which indicates that only for $B_{0}\gtrsim m$ (when $\vert\vec{B}\vert$ is very 
small, as~could be in a certain spacetime, e.g.,~Bianchi II universe with
equal scale factors, shown above), the~gravitational field 
could affect the oscillation, which is again in accordance with 
experiments done in the laboratory~\cite{exp}. However, this condition 
could easily be satisfied in few factors to few tens of Schwarzschild radii away 
from a primordial black hole of mass $\lesssim$$10^{24}$--$10^{20}$ gm.

\subsection{\label{flmx}Flavor~Mixing}

As each flavor acquires two mass states through particle--antiparticle
mixing, a~two-generation electron--muon neutrino system effectively will have four flavor states,
governing the Lagrangian density~\cite{bm2}
\begin{eqnarray} \nonumber (-g)^{-1/2} {\cal
L}_m~&=&~-\frac12\left(\nu_{e1}^\dag
m_{e1}\nu_{e1}~+~\nu_{e2}^\dag m_{e2}\nu_{e2}~+~\nu_{\mu1}^\dag
m_{\mu1}\nu_{\mu1}~+~\nu_{\mu2}^\dag
m_{\mu2}\nu_{\mu2}\right.\\&-& \left.~\nu_{\mu1}^\dag m_{\mu e}
\nu_{e1}~-~\nu_{e1}^\dag m_{\mu e}\nu_{\mu 1}~+~\nu_{\mu2}^\dag
m_{\mu e} \nu_{e2}~+~\nu_{e2}^\dag m_{\mu e}
\nu_{\mu2}\right),
\label{lagmass}
\end{eqnarray}
in the presence of a mixing mass $m_{\mu e}$.
From Equation~(\ref{mas})
\begin{eqnarray}
m_{(e,\mu)1}~=~m_n-\sqrt{B_0^2+m_{e, \mu}^2},\,\,\,\,\, 
m_{(e,\mu)2}~=~m_n+\sqrt{B_0^2+m_{e, \mu}^2},
\label{flvmas}
\end{eqnarray}
which no longer are the definite masses of electron and muon neutrinos due to the presence of $m_{\mu e}$.
In terms of mass states $|\chi_i\rangle$, when $i=1-4$, the~sets of 
two Majorana neutrino flavor eigenstates are described as
\begin{subequations}\label{fl1}\begin{eqnarray} |\nu_{e1}
\rangle~=~\cos{\theta_1}|\chi_1\rangle~-~e^{i\phi_1}~\sin{\theta_1}|\chi_2\rangle \\
|\nu_{\mu1}\rangle~=~\sin{\theta_1}|\chi_1\rangle~+~e^{i\phi_1}~\cos{\theta_1}|\chi_2\rangle
\end{eqnarray}\end{subequations} and \begin{subequations}\label{fl2}\begin{eqnarray} |\nu_{e2}
\rangle~=~\cos{\theta_2}|\chi_3\rangle~-~e^{i\phi_2}~\sin{\theta_2}|\chi_4\rangle \\
|\nu_{\mu2}\rangle~=~\sin{\theta_2}|\chi_3\rangle~+~e^{i\phi_2}~\cos{\theta_2}|\chi_4\rangle,
\end{eqnarray}\end{subequations}
where the corresponding mixing parameters are given by
\begin{equation} \tan \theta_{1,2}~=~\frac{\mp 2m_{\mu
e}}{(m_{e(1,2)}-m_{\mu(1,2)})+\sqrt{(m_{e(1,2)}-m_{\mu(1,2)})^2+
4m_{\mu e}^2}},\,\,\,\,\,\,\phi_{1,2}=arg(\mp m_{\mu e}). \label{tant}
\end{equation}

At an arbitrary time $t$, $|\chi_{1,3}(0)\rangle\rightarrow|\chi_{1,3}(t)
\rangle=e^{-iE_{e(1,2)}t}|\chi_{1,3}(0)\rangle$ and 
$|\chi_{2,4}(0)\rangle\rightarrow|\chi_{2,4}(t)
\rangle=e^{-iE_{\mu(1,2)}t}|\chi_{2,4}(0)\rangle$, when $E_{e(1,2)}=
\sqrt{|\vec{p}|^2+m_{e(1,2)}^2}$ and
$E_{\mu(1,2)}=\sqrt{|\vec{p}|^2+m_{\mu(1,2)}^2}$.

The oscillation lengths can be recalled as~\cite{bm2}
\begin{equation} \lambda_{fg1}~=~ \frac{4\pi E}{|M_1^2-M_2^2|},\,\,\,\,
\lambda_{fg2}~=~ \frac{4\pi E}{|M_3^2-M_4^2|},
\end{equation}
where
\begin{eqnarray}
\Delta M^2=|M_{1,3}^2-M_{2,4}^2| \nonumber \\
=~\left(\sqrt{B_0^2+m_{\mu(1,2)}^2}+\sqrt{B_0^2+m_{e(1,2)}^2}\right) \nonumber
\\
\times\sqrt{\left\{\left(\sqrt{B_0^2+m_{\mu(1,2)}^2}-\sqrt{B_0^2+m_{e(1,2)}^2}\right)^2+4m_{e\mu}^2\right\}},
\end{eqnarray}
which indicates that only for $B_0\gtrsim m_{e(1,2)},m_{\mu(1,2)}$, the~gravitational field 
could affect the oscillation. For~$B_0>> m$, mixing is maximum 
{{and the oscillation 
length turns out to be} $\lambda_{fg1,2}\sim\pi E/B_0m_{e\mu}$. {For~other 
quantitative details see}~\cite{bm2}.}

In the more realistic three-flavor case, the~above discussions remain valid,
but with the emergence of three sets of mass eigenstates. All the underlying masses will also
be modified by the gravitational coupling term, similar to those given by
Equations~(\ref{mas}) and (\ref{flvmas}). Accordingly, mixing angle will become complicated and also
the oscillation probability, but~the influence of gravitational field will remain there,
depending on, e.g.,~the black hole~mass.

\section{Dynamic and Geometric~Phases}\label{sec4}

Let us consider the wavefunction $\Psi(t)$ of a system evolving over a time interval $t \in [0,\tau]$,
where $\Psi(0)$ is its initial state and $\Psi(\tau)$ being the final. The~total phase accumulated 
over the entire evolution is given by $\Phi_t=arg(\langle\Psi(0)\vert\Psi(\tau) \rangle)$ and the 
corresponding dynamic phase is given by 
$\Phi_d=-\int_{0}^{\tau}\langle\Psi(t)\vert i\partial_{t}\vert\Psi(t) \rangle dt$. The~difference between
the two phases is defined as the geometric phase~\cite{muksi}, given by
\begin{equation} 
\Phi_g= arg(\langle\Psi(0)\vert\Psi(\tau) \rangle) + \int_{0}^{\tau} \langle\Psi(t)\vert i\partial_{t}\vert\Psi(t) \rangle dt.
\end{equation} 

In a situation where the system could oscillate back-and-forth between $\Psi(t)$ and 
$\tilde{\Psi}(t)$ (which is antiparticle of $\Psi(t)$), e.g., the case of neutrino oscillation, we define new total and dynamic phases respectively given by
$\Phi_{to}=arg(\langle\tilde{\Psi}(0)\vert\Psi(\tau) \rangle)$ and 
$\Phi_{do}=-\int_{0}^{\tau}\langle\tilde{\Psi}(t)\vert i\partial_{t}\vert\Psi(t) \rangle dt$.
We term them as respective oscillation phases, where the geometric oscillation phase is given by
\begin{equation} 
\Phi_{go} = arg(\langle\tilde{\Psi}(0)\vert\Psi(\tau) \rangle) + \int_{0}^{\tau} \langle\tilde{\Psi}(t)\vert i\partial_{t}\vert\Psi(t) \rangle dt.
\end{equation}

Below we use these definitions to evaluate various phases in the neutrino sector.
More precisely, we evaluate $\Phi_t$, $\Phi_d$, $\Phi_g$ and $\Phi_{to}$, 
$\Phi_{do}$, $\Phi_{go}$ for various
neutrino states recalled in the previous~section.

The phases, as~we show below, depend on $B_\mu$, which furthermore is determined by
the nature of underlying spacetime and the corresponding parameter values. For~the 
explicit computations of $B_\mu$, see previous papers, e.g.,~\cite{bm1,bm2,bmmpla,ujjal,
param,bm02}. Nevertheless, for~the present purpose we do not consider the contribution 
due to the spatial variation of neutrino states at $t=0$, which is obvious from 
{Section} 
 \ref{sec2}. Our interest rather is the contribution to the geometric phases due to 
mixing and oscillation of states, which arise due to the effect of spacetime curvature 
on to the time evolution of neutrino~states. 

\subsection{Neutrino-Antineutrino~Mixing}\label{sec.4.1}

The total phase for $|\nu_1\rangle$ is
\begin{eqnarray}
\Phi_{t1}=arg\left(\langle\nu_1(0)|\nu_1(\tau)\rangle\right)
=\tan^{-1}\left(-\frac{\cos^2\theta\sin E_\psi \tau+\sin^2\theta\sin E_{\psi^c}\tau}{\cos^2\theta\cos E_{\psi}\tau
+\sin^2\theta\cos E_{\psi^c}\tau}\right)
\end{eqnarray}
and for $|\nu_2\rangle$
\begin{eqnarray}
\Phi_{t2}=arg\left(\langle\nu_2(0)|\nu_2(\tau)\rangle\right)
=\tan^{-1}\left(-\frac{\sin^2\theta\sin E_\psi \tau+\cos^2\theta\sin E_{\psi^c}\tau}{\sin^2\theta\cos E_{\psi}\tau
+\cos^2\theta\cos E_{\psi^c}\tau}\right).
\end{eqnarray}

The other phase
is given by for $|\nu_1\rangle$ as
\begin{eqnarray}
-\Phi_{d1}=\int_0^\tau\langle\nu_1(t)|i\partial_t |\nu_1(t)\rangle dt=-\phi~\sin^2\theta+\int_0^\tau\left(E_\psi\cos^2\theta+
E_{\psi^c}\sin^2\theta\right)dt
\label{g1}
\end{eqnarray}
and for $|\nu_2\rangle$ as
\begin{eqnarray}
-\Phi_{d2}=\int_0^\tau\langle\nu_2(t)|i\partial_t |\nu_2(t)\rangle dt=-\phi~\cos^2\theta+\int_0^\tau\left(E_\psi\sin^2\theta+
E_{\psi^c}\cos^2\theta\right)dt,
\label{g2}
\end{eqnarray}
when $\theta$ is independent of time. 
Even if $\theta$ is not constant, 
the part outside the integral in either of the Equations~(\ref{g1}) and (\ref{g2}) always contributes
to $\Phi_{g1}$ and $\Phi_{g2}$ respectively, revealing a $\tau$-independent phase, as~long as $\phi$
in the neutrino states is not constant.
For $\theta=0$ which corresponds to $B_0>>m$, the~total geometric phases for $\nu_1$ and $\nu_2$ 
turn out to be $n\pi$ and $n\pi-\phi$ respectively with $n=0,1,2,3\cdots$. For~$\theta=\pi/4$ which 
corresponds to $B_0<<m$, they are $n\pi-\phi/2$. However, generally speaking neutrino mass does
not vary with time and hence $\phi$ remains fixed throughout the propagation. Thus,~all the terms
associated with $\phi$ actually vanish and any geometric contribution to the phase arises
from other terms in, e.g.,~$\Phi_{t1}-\Phi_{d1}$. 

For the oscillation between mass eigenstates, total phase
\begin{eqnarray}
\Phi_{to1}=arg\left(\langle\nu_2(0)|\nu_1(\tau)\rangle\right)
=\frac{\pi}{2}-\frac{(E_\psi+E_{\psi^c})\tau}{2}
\label{to1}
\end{eqnarray}
and the other phase 
is given by
\begin{eqnarray}
-\Phi_{do1}=\int_0^\tau\langle\nu_2(t)|i\partial_t |\nu_1(t)\rangle dt=\phi~
\cos\theta\sin\theta+\int_0^\tau\left(E_\psi-E_{\psi^c}\right)\cos\theta\sin\theta~dt.
\label{go1}
\end{eqnarray}

As before, the~term outside the integral in $\Phi_{do1}$ survives only if $\phi$ varies with time,
which generally may not be the case as the neutrino mass is~fixed.

From previous work~\cite{bmmpla,bm1}, $B_0$ can be computed for the spacetime around black holes as
\begin{eqnarray}
B_0\sim 10^{-18}\frac{M_\odot}{M}{\rm GeV}.
\end{eqnarray}

Therefore, for~a
black hole in an X-ray binary with $M=10M_\odot$, $B_0<<m$, when the Majorana mass of a neutrino $m\sim 10^{-2}$eV. 
In this case, there is apparently no effect of gravity on the geometric and dynamic phases
and $\Phi_{do1}$ turns out to be $\tau$-independent and arises due to the Majorana nature of the neutrino
because $E_\psi=E_\psi^c$. 
This is purely the consequence of the mixing of neutrino and antineutrino, which occurs due to the 
presence of Majorana mass. The~same is true for black holes at the center of~AGNs.

For primordial black holes with $M\le 10^{24}$ gm, on~the other hand, $B_0\ge 1$ eV so that $B_0>>m$.
Therefore, $\theta\rightarrow 0$ and hence the part outside the integral of $\Phi_{d1}\rightarrow 0$ and and that of $\Phi_{d2}\rightarrow -\phi$. Moreover, 
$\Phi_{do1}\rightarrow 0$. In~this case, gravitational
field removes any possibility of mixing and then oscillation, which 
however affects geometric and dynamic~phases.

When the mass of a primordial black hole increases to $M=10^{26}$ gm, $B_0\sim m$, which alters
the mixing angle compared to that in the absence of gravitational effect, and~hence affects the phases.
An important point to note is that the larger the mass of black hole, the larger  
its radius, and~hence the smaller the density in the surrounding disk. 
Therefore, in~order to affect geometric and dynamic phases due to gravitational effect, 
the gravitational mass should not be
more than $\sim$$10^{-6}M_\odot$. 

\subsection{Mixing of Mass~Eigenstates}\label{sec.4.2}

In this case, for~$|\psi^c\rangle$ the total phase
\begin{eqnarray}
\Phi_{t1}=arg\left(\langle\psi^c(0)|\psi^c(\tau)\rangle\right)
=\tan^{-1}\left(-\frac{\cos^2\theta\sin E_2 \tau+\sin^2\theta\sin E_1 \tau}{\cos^2\theta\cos E_2 \tau
+\sin^2\theta\cos E_1 \tau}\right)
\end{eqnarray}
and the dynamical phase 
containing a term which does not explicitly depend on $\tau$ due to non-zero neutrino phase $\phi$, given by
\begin{eqnarray}
-\Phi_{d1}=\int_0^\tau\langle\psi^c(t)|i\partial_t |\psi^c(t)\rangle dt=-\phi~\sin^2\theta+\int_0^\tau\left(E_2\cos^2\theta+
E_{1}\sin^2\theta\right)dt.
\label{g11}
\end{eqnarray}

Similarly, for~$|\psi\rangle$
\begin{eqnarray}
\Phi_{t2}=arg\left(\langle\psi(0)|\psi(\tau)\rangle\right)
=\tan^{-1}\left(-\frac{\sin^2\theta\sin E_2 \tau+\cos^2\theta\sin E_{1}\tau}{\sin^2\theta\cos E_{2}\tau
+\cos^2\theta\cos E_{1}\tau}\right),
\end{eqnarray}
\begin{eqnarray}
-\Phi_{d2}=\int_0^\tau\langle\psi(t)|i\partial_t |\psi(t)\rangle dt=-\phi~\cos^2\theta+\int_0^\tau\left(E_2\sin^2\theta+
E_{1}\cos^2\theta\right)dt.
\label{g22}
\end{eqnarray}

For neutrino-antineutrino oscillation, the~total phase
\begin{eqnarray}
\Phi_{to1}=arg\left(\langle\psi^c(0)|\psi(\tau)\rangle\right)
=\frac{\pi}{2}-\frac{(E_1+E_2)\tau}{2}
\label{to1}
\end{eqnarray}
and the other phase 
\begin{eqnarray}
-\Phi_{do1}=\int_0^\tau\langle\psi^c(t)|i\partial_t |\psi(t)\rangle dt=\phi~
\cos\theta\sin\theta+\int_0^\tau\left(E_1-E_2\right)\cos\theta\sin\theta~dt.
\label{go1}
\end{eqnarray}

Here $\theta$ is assumed to be independent of time. If, in~general, $\theta$ is not a 
constant, then other terms will contribute to $\Phi_{d1,2}$ and $\Phi_{do1}$.
For $B_0>>m$, $\Phi_{do1}\rightarrow 0$, 
while for $m>>B_0$, all parts of the phases survive. Oscillation is also possible 
for $m>>B_0$, as~long as $m_n\neq 0$.
Simultaneously, oscillation and modified 
geometric and dynamic phases due to gravity are revealed, only when $B_0\sim m,m_n$, which is possible in the site of, e.g.,~primordial black holes.
Note that $\theta$ and $\phi$ are the same as that for the cases of neutrino-antineutrino mixing for the various parameters of spacetime, e.g.,
the mass of the black hole. More so, as~mentioned before, any term in the phase associated with $\phi$ does not
survive if $\phi$ is not a time-varying function, which generally is the case for neutrinos whose mass
is assumed to be~fixed.

\subsection{Flavor~Mixing}

In this case, the~total phase for $|\nu_{e1,2}\rangle$
\begin{eqnarray}
	\nonumber
\Phi_{t1}=arg\left(\langle\nu_{e(1,2)}(0)|\nu_{e(1,2)}(\tau)\rangle\right)
=\tan^{-1}\left(-\frac{\cos^2\theta_{1,2}\sin E_{e(1,2)} \tau+\sin^2\theta_{1,2}\sin E_{\mu(1,2)} \tau}
{\cos^2\theta_{1,2}\cos E_{e(1,2)} \tau
+\sin^2\theta_{1,2}\cos E_{\mu(1,2)} \tau}\right)\\
\end{eqnarray}
and the other phase containing a $\tau$-independent part, given by
\begin{eqnarray}
\nonumber
-\Phi_{d1}=\int_0^\tau\langle\nu_{e(1,2)}(t)|i\partial_t |\nu_{e(1,2)}(t)\rangle dt=-\phi_{1,2}~\sin^2\theta_{1,2}+\int_0^\tau\left(E_{e(1,2)}\cos^2\theta_{1,2}+
E_{\mu(1,2)}\sin^2\theta_{1,2}\right)dt.\\
\label{g111}
\end{eqnarray}

For $|\nu_{\mu(1,2)}\rangle$, they are
\begin{eqnarray}
\nonumber
\Phi_{t2}=arg\left(\langle\nu_{\mu(1,2)}(0)|\nu_{\mu(1,2)}(\tau)\rangle\right)
=\tan^{-1}\left(-\frac{\sin^2\theta_{1,2}\sin E_{e(1,2)} \tau+\cos^2\theta_{1,2}\sin E_{\mu(1,2)}\tau}
{\sin^2\theta_{1,2}\cos E_{e(1,2)}\tau +\cos^2\theta_{1,2}\cos E_{\mu(1,2)}\tau}\right),\\
\end{eqnarray}
\begin{eqnarray}
\nonumber
-\Phi_{d2}=\int_0^\tau\langle\nu_{\mu(1,2)}(t)|i\partial_t |\nu_{\mu(1,2)}(t)\rangle dt=-\phi_{1,2}~\cos^2\theta_{1,2}+
\int_0^\tau\left(E_{e(1,2)}\sin^2\theta_{1,2}+ E_{\mu(1,2)}\cos^2\theta_{1,2}\right)dt,\\
\label{g222}
\end{eqnarray}
when $\theta_{1,2}$ are independent of time. As~before, for~$\theta_{1,2}$ not
being constant, other complicated dynamical terms will contribute to 
$\Phi_{g1,2}$, and~the phases associated with $\phi$ would not contribute eventually as the neutrino mass
does not vary with time in~general.

For oscillation, total phase
\begin{eqnarray}
\Phi_{to1}=arg\left(\langle\nu_{\mu(1,2)}(0)|\nu_{e(1,2)}(\tau)\rangle\right)
=\frac{\pi}{2}-\frac{(E_{e(1,2)}+E_{\mu(1,2)})\tau}{2}
\label{to22}
\end{eqnarray}
and other phase
\begin{eqnarray}
	\nonumber
-\Phi_{do1}=\int_0^\tau\langle\nu_{\mu(1,2)}(t)|i\partial_t |\nu_{e(1,2)}(t)\rangle dt
=\phi_{1,2}\cos\theta_{1,2}\sin\theta_{1,2}+\int_0^\tau(E_{e(1,2)}-E_{\mu(1,2)})
\cos\theta_{1,2}\sin\theta_{1,2}~dt,\\
\label{go22}
\end{eqnarray}
where $E_{e(1,2)}-E_{\mu(1,2)}=\Delta M^2/2|\vec{p}|$ and 
$E_{e(1,2)}+E_{\mu(1,2)}=2|\vec{p}|+(m_{e(1,2)}^2+m_{\mu(1,2)}^2)/2|\vec{p}|$ in the weak gravity limit.
When $B_0>>m_n,m_{e(1,2)},m_{\mu(1,2)}$, interestingly $\theta_{1,2}$ given in Equation~(\ref{tant}) become constant and equal to $\mp\pi/4$ (when the lepton number
conserving mass $m_n$ is the same in both the electron and muon sectors).
This brings the part outside the integrals in $\Phi_{d1,2}$ and $\Phi_{do1}$ as 
a constant which furthermore turns out to be the same as the geometric phase 
$\Phi_{g1,2}$. They are 
independent of whether the spacetime is stationary
(e.g., around a rotating black hole) or time-dependent (e.g., of early universe). Note that in the absence of 
gravity ($B_0<<m_n,m_{e1,2},m_{\mu1,2}$), $\theta_{1,2}$ depend on specific values of neutrino masses.
However, if~$m_{e1,2}=m_{\mu1,2}$, then again $\theta_{1,2}=\mp\pi/4$ 
(when the lepton number conserving mass $m_n$ is the same in both the electron and muon sectors). Also, the dynamical parts of flavor oscillation phase survive
whether $B_0$ dominating neutrino masses or otherway round, however
their variation depends on the value of $B_0$.

\section{Entanglement of Neutrino States and Corresponding Geometric~Phases}\label{sec5}

We begin by showing that neutrino ($\psi$) and antineutrino ($\psi^c$) 
combined system, as~given by Equation~(\ref{eqn21}) (also Equation~(\ref{mix})), forms entanglement. 
As $\psi^c=-i\sigma_2\psi^*$, and~if $\psi^c$ is purely spin-down with only one component non-zero
then $\psi$ is purely spin-up,
\begin{equation} \left(
   \begin{array}{cc}
    \psi^c \\
    \psi \\ \end{array}
   \right)=
c\left(
   \begin{array}{cc}
    0 \\
    1 \\ \end{array}
   \right)\otimes
\left(
   \begin{array}{cc}
    1 \\
    0 \\ \end{array}
   \right)-
c^*\left(
   \begin{array}{cc}
    1 \\
    0 \\ \end{array}
   \right)\otimes
\left(
   \begin{array}{cc}
    0 \\
    1 \\ \end{array}
   \right),
    \end{equation}
where $c$ is the non-zero component of $\psi$ and we choose Weyl representation
for convenience. As~it stands, the~first and second terms cannot be 
decomposed into the direct-product of two independent states, hence they
entangle. Similarly, the combined 
mass eigenstates in the presence of gravitational field and Majorana mass, 
given by Equation~(\ref{mix}), can be shown to exhibit entangled~states.

Now in the presence of flavor mixing, as~given by 
Equations~(\ref{fl1}) and (\ref{fl2}), the~states $\nu_{e1}$ and $\nu_{\mu1}$
are orthogonal to each other and the states $\nu_{e2}$ and $\nu_{\mu2}$ do so.
Also without mixing term, $\nu_{e1}$ and $\nu_{e2}$ form two orthogonal
mass eigenstates for neutrino--antineutrino mixing in the electron sector
and $\nu_{\mu1}$ and $\nu_{\mu2}$ in the muon sector (when we consider
only two flavors for simplicity).

Interestingly, it is clear from Equations~(\ref{mixt}) and (\ref{energy}) that 
gravitational field converts $\nu_{e1}$ (and~$\nu_{\mu1}$) to $\nu_{e2}$ (and~$\nu_{\mu2}$) 
by oscillation, leading to both of them being present at an arbitrary time. Hence,~Equations~(\ref{fl1}) and (\ref{fl2}) 
show that gravitational effect brings out two independent sets of flavor
neutrinos, $\{\nu_{e1},\nu_{\mu1}\}$ and $\{\nu_{e2},\nu_{\mu2}\}$,
satisfying respective orthogonality conditions between electron and
muon neutrinos in the respective Hilbert spaces ${\cal H}_1$ and
${\cal H}_2$ independently. Hence,
the neutrino states in ${\cal H}_1$ should entangle with those in ${\cal H}_2$ which are non-interacting. 
Therefore, following the conventional approach (e.g.,~\cite{eriks}) we can construct
the entangled states at $t=0$
\begin{subequations}
\label{ent}
\begin{eqnarray}
|\psi_1(0)\rangle=\cos\alpha\,|\nu_{e1}(0)\rangle |\nu_{e2}(0)\rangle+
e^{i\beta}\,\sin\alpha\,|\nu_{\mu1}(0)\rangle |\nu_{\mu2}(0)\rangle,\\
|\psi_2(0)\rangle=-\sin\alpha\,|\nu_{e1}(0)\rangle |\nu_{e2}(0)\rangle+
e^{i\beta}\,\cos\alpha\,|\nu_{\mu1}(0)\rangle |\nu_{\mu2}(0)\rangle,\\
|\psi_3(0)\rangle=\cos\alpha\,|\nu_{e1}(0)\rangle |\nu_{\mu2}(0)\rangle+
e^{i\beta}\,\sin\alpha\,|\nu_{\mu1}(0)\rangle |\nu_{e2}(0)\rangle,\\
|\psi_4(0)\rangle=-\sin\alpha\,|\nu_{e1}(0)\rangle |\nu_{\mu2}(0)\rangle+
e^{i\beta}\,\cos\alpha\,|\nu_{\mu1}(0)\rangle |\nu_{e2}(0)\rangle,
\end{eqnarray}
\end{subequations}
when $|\nu_{e1}\rangle$ and $|\nu_{e2}\rangle$ 
(and $|\nu_{\mu1}\rangle$ and $|\nu_{\mu2}\rangle$) in Equation~(\ref{ent}a) are 
two points on the Poincar\'e sphere and so on for others equations. The~angle $\alpha$ 
determines the degree of entanglement. As~is the case in the Poincar\'e sphere
of a single spin$-$1/2 particle, the above~equation suggests that 
$\alpha$ and $\beta$ parameterize a two-sphere called Schmidt~sphere.

{{Note that various quantum correlations of the system of basic 
one-flavor and two-flavor neutrino and antineutrino states 
in the presence of gravitational field and hence gravitational Zeeman splitting, as~described in Section \ref{solution}, were already explored by one 
of the present authors}~\cite{khushbooBM}.
{Flavor entropy has been used therein to probe the entanglement in the system, which 
gets suppressed with the increase of gravitational field.} }

At an arbitrary time $t$, the~entangled states, {{defined in
Equation}~(\ref{ent}a--d)}, go to $|\psi_{1,2,3,4}(t)\rangle$ which have the
same form as in Equation~(\ref{ent}a--d), except~$|\nu_{e1,2}(0)\rangle$ and $|\nu_{\mu1,2}(0)\rangle$
replaced by $|\nu_{e1,2}(t)\rangle$ and $|\nu_{\mu1,2}(t)\rangle$ respectively, as~given by
Equations~(\ref{fl1}) and (\ref{fl2}) in terms of $|\chi_{1,3}(t)\rangle$ and $|\chi_{2,4}(t)\rangle$ generically.

Therefore, based on the definitions given in the beginning of Section \ref{sec4}, the~total,
dynamic and geometric parts 
of the phases in the evolution of entangled states $\psi_{1,2,3,4}$ can be obtained from
\begin{eqnarray}
\Phi_{t1,2,3,4}=arg\left(\langle\psi_{1,2,3,4}(0)|\psi_{1,2,3,4}(\tau)\rangle\right),~~~
-\Phi_{d1,2,3,4}=\int_0^\tau\langle\psi_{1,2,3,4}(t)|i\partial_t|\psi_{1,2,3,4}(t)\rangle dt.
\end{eqnarray}

Explicitly, for~$\alpha=\pi/4$ and $\beta=0$, we obtain
\begin{equation}
\begin{array}{l}
\Phi_{t1,2_{\pi/4}}=\tan^{-1}\\
\left[-\frac{\cos^2\theta^\mp_{12}\{\sin(E_{e1}+E_{e2})\tau+
\sin(E_{\mu1}+E_{\mu2})\tau\}+
\sin^2\theta^\mp_{12}\{\sin(E_{\mu1}+E_{e2})\tau+
\sin(E_{\mu2}+E_{e1})\tau\}}
{\cos^2\theta^\mp_{12}\{\cos(E_{e1}+E_{e2})\tau+
\cos(E_{\mu1}+E_{\mu2})\tau\}+\sin^2\theta^\mp_{12}\{\cos(E_{\mu1}+E_{e2})\tau+
\cos(E_{\mu2}+E_{e1})\tau\}}
\right],\\
\end{array}
\end{equation}
\begin{equation}
\begin{array}{l}
\Phi_{t3,4_{\pi/4}}=\tan^{-1}\\
\left[-\frac{\sin^2\theta^\pm_{12}\{\sin(E_{e1}+E_{e2})\tau+
\sin(E_{\mu1}+E_{\mu2})\tau\}+
\cos^2\theta^\pm_{12}\{\sin(E_{\mu1}+E_{e2})\tau+
\sin(E_{\mu2}+E_{e1})\tau\}}
{\sin^2\theta^\pm_{12}\{\cos(E_{e1}+E_{e2})\tau+
\cos(E_{\mu1}+E_{\mu2})\tau\}+\cos^2\theta^\pm_{12}\{\cos(E_{\mu1}+E_{e2})\tau+
\cos(E_{\mu2}+E_{e1})\tau\}}
\right],\\
\end{array}
\end{equation}
when $\cos^2\theta^\mp_{12}=\cos^2(\theta_1\mp\theta_2)$
and $\sin^2\theta^\mp_{12}=\sin^2(\theta_1\mp\theta_2)$.
For $\theta_1=\theta_2$ and $\theta_1=-\theta_2$, respectively for $\Phi_{t1,4_{\pi/4}}$ and 
$\Phi_{t2,3_{\pi/4}}$, the~phases furthermore reduce to
\begin{eqnarray}
\Phi_{t1,2,3,4_{\pi/4}}=-\left(E_{e1}+E_{e2}+E_{\mu1}+E_{\mu2}\right)\frac{\tau}{2}~~~{\rm or}~~~
\pi-\left(E_{e1}+E_{e2}+E_{\mu1}+E_{\mu2}\right)\frac{\tau}{2}.
\end{eqnarray}

For constant and arbitrary $\alpha,\theta_1,\theta_2$
\begin{eqnarray}
\begin{array}{l}
\label{entber}
-\Phi_{d1}=\\
\cos^2\alpha\left[-\phi_1\sin^2\theta_1-\phi_2\sin^2\theta_2+\int_0^\tau\left(E_{\mu1}\cos^2\theta_1+E_{e1}\sin^2\theta_1+E_{\mu2}\cos^2\theta_2
+E_{e2}\sin^2\theta_2\right)dt\right]+\\
\sin^2\alpha\left[-\phi_1\cos^2\theta_1-\phi_2\cos^2\theta_2+\int_0^\tau\left(E_{\mu1}\sin^2\theta_1+E_{e1}\cos^2\theta_1+E_{\mu2}\sin^2\theta_2
+E_{e2}\cos^2\theta_2\right)dt\right]\\
-\beta\sin^2\alpha
\end{array}
\end{eqnarray}
and $-\Phi_{d2}$ is the same except $\cos^2\alpha$ and $\sin^2\alpha$ are interchanged.
Similarly,
\begin{eqnarray}
\begin{array}{l}
\label{entber}
-\Phi_{d3}=\\
\cos^2\alpha\left[-\phi_1\sin^2\theta_1-\phi_2\cos^2\theta_2+\int_0^\tau\left(E_{\mu1}\sin^2\theta_1+E_{e1}\cos^2\theta_1+E_{\mu2}\cos^2\theta_2
+E_{e2}\sin^2\theta_2\right)dt\right]+\\
\sin^2\alpha\left[-\phi_1\cos^2\theta_1-\phi_2\sin^2\theta_2+\int_0^\tau\left(E_{\mu1}\cos^2\theta_1+E_{e1}\sin^2\theta_1+E_{\mu2}\sin^2\theta_2
+E_{e2}\cos^2\theta_2\right)dt\right]\\
-\beta\sin^2\alpha
\end{array}
\end{eqnarray}
and $-\Phi_{d4}$ is the same except $\cos^2\alpha$ and $\sin^2\alpha$ are~interchanged.

For $\alpha=\pi/4$, $\Phi_{d1,2,3,4}$ reduce as
\begin{eqnarray}
-\Phi_{d1,2,3,4_{\pi/4}}=\frac{1}{2}\left[-\phi_1-\phi_2+\int_0^\tau\left(E_{e1}+E_{e2}+E_{\mu1}+E_{\mu2}\right)dt-
\beta\right].
\end{eqnarray}

For oscillation between entangled states, the~total phases 
$\Phi_{to1}=arg(\langle\psi_2(0)|\psi_1(\tau)\rangle)$ and 
$\Phi_{to2}=arg(\langle\psi_4(0)|\psi_3(\tau)\rangle)$, for~$\alpha=\pi/4$ and $\beta=0$, are
\begin{eqnarray}
\Phi_{{to1,2}_{\pi/4}}=\frac{\pi}{2}-\left(E_{e1}+E_{e2}+E_{\mu1}+E_{\mu2}\right)\frac{\tau}{2}.
\end{eqnarray}

The other phases, containing 
a $\tau$-independent part during oscillation, 
$-\Phi_{do1}=\int_0^\tau\langle\psi_2(t)|i\partial_t|\psi_1(t)\rangle dt$  and 
$-\Phi_{do2}=\int_0^\tau\langle\psi_4(t)|i\partial_t|\psi_3(t)\rangle dt$, are given by
\begin{eqnarray}
\begin{array}{l}
\label{entober}
-\Phi_{do1,2}=\\
-\cos\alpha\sin\alpha\left[\cos2\theta_1\left\{\phi_1+\int_0^\tau\left(E_{e1}-E_{\mu1}\right)dt\right\}\pm
\cos2\theta_2\left\{\phi_2+\int_0^\tau\left(E_{e2}-E_{\mu2}\right)dt\right\}+\beta\right].
\end{array}
\end{eqnarray}

For varying mixing parameters, the~terms appearing outside the integral 
in all $\Phi_d$-s will also always contribute to the respective phases. 
Note interestingly that
\begin{eqnarray}
\Phi_{d1}+\Phi_{d2}=\Phi_{d3}+\Phi_{d4}=
-\left(E_{e1}+E_{\mu1}+E_{e2}+E_{\mu2}\right)\tau+\left(\beta+\phi_1
+\phi_2\right).
\end{eqnarray}

Like the cases in previous section, as~neutrino mass is not expected to vary with time,
the phases associated with $\phi$ do not contribute~generally.

\section{Variation of Mixing Angles with Gravitational~Field}\label{sec6}

\begin{figure}
        \centering
        \fbox{\includegraphics[width=0.9\linewidth]{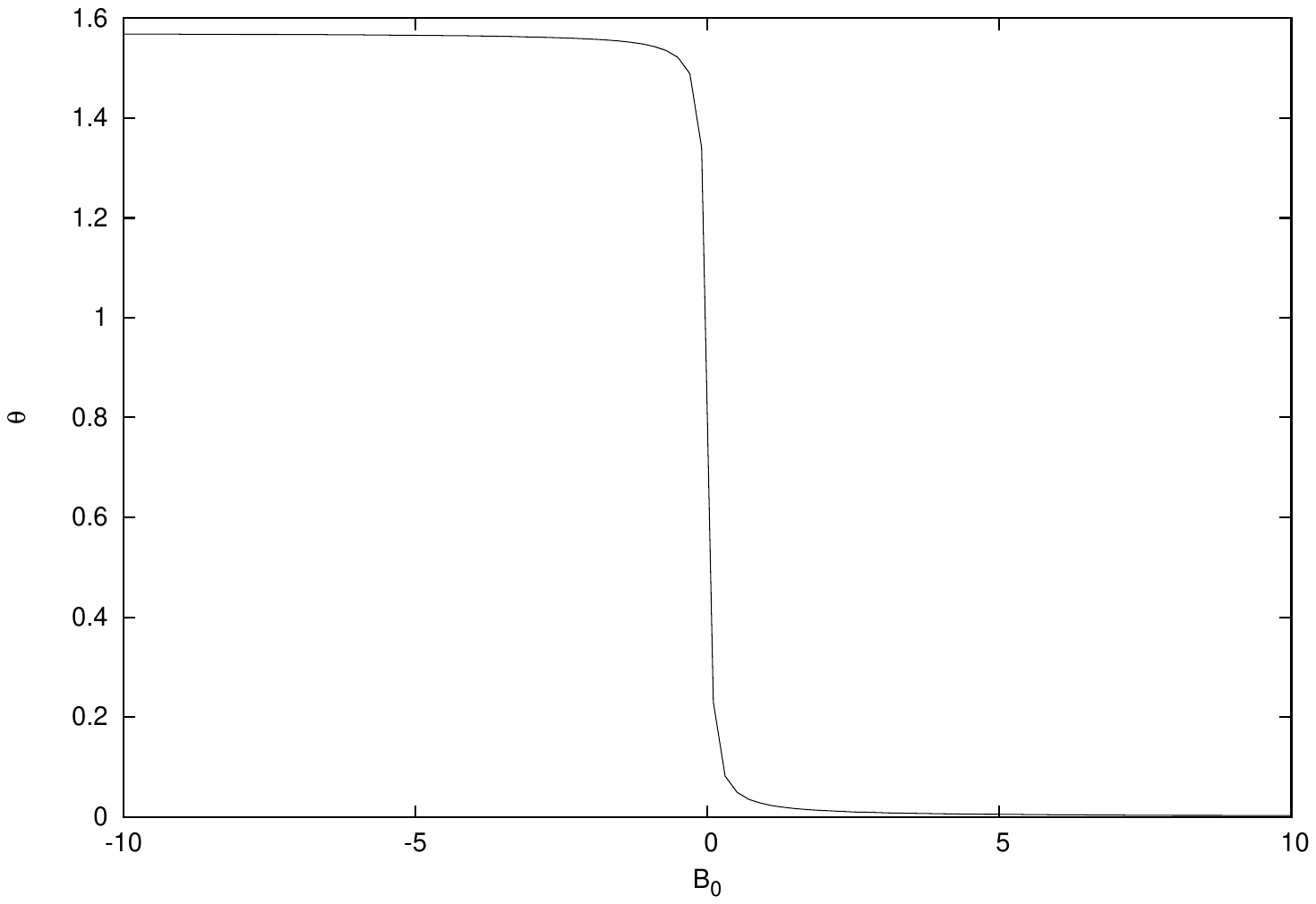}}
        \caption{
Variation of mixing angle in radian of basic neutrino--antineutrino mixing as a function of 
gravitational coupling in units of eV with $m=0.05$ eV.
}
        \label{figth}
\end{figure}

The phases independent of $\tau$ are associated with mixing angles and phases of neutrinos, e.g.,
$\theta_1, \theta_2, \alpha$ (also $\theta$) and $\phi_1, \phi_2, \beta$ (also $\phi$).
Therefore, depending on $\theta$-s, which are determined by gravitational field and the physical nature of spacetime geometry, 
the $\tau$-independent parts of phases vary. 
In the absence of gravitational field and in the presence of lepton number violating 
interaction and hence Majorana mass, neutrino and antineutrino mix with $\theta=\pi/4$.
Figure~\ref{figth} shows that how the mixing angle of the basic neutrino-antineutrino states
changes with gravitational coupling, which~furthermore controls the geometric Berry-like 
phases associated with $\Phi_{t1,2}-\Phi_{d1,2}$ given in Sections \ref{sec.4.1} and \ref{sec.4.2}.
While a stronger gravity effect kills oscillation, it still leads to a
non-zero geometric~phase.

\begin{figure}
        \centering
        \fbox{\includegraphics[width=0.9\linewidth]{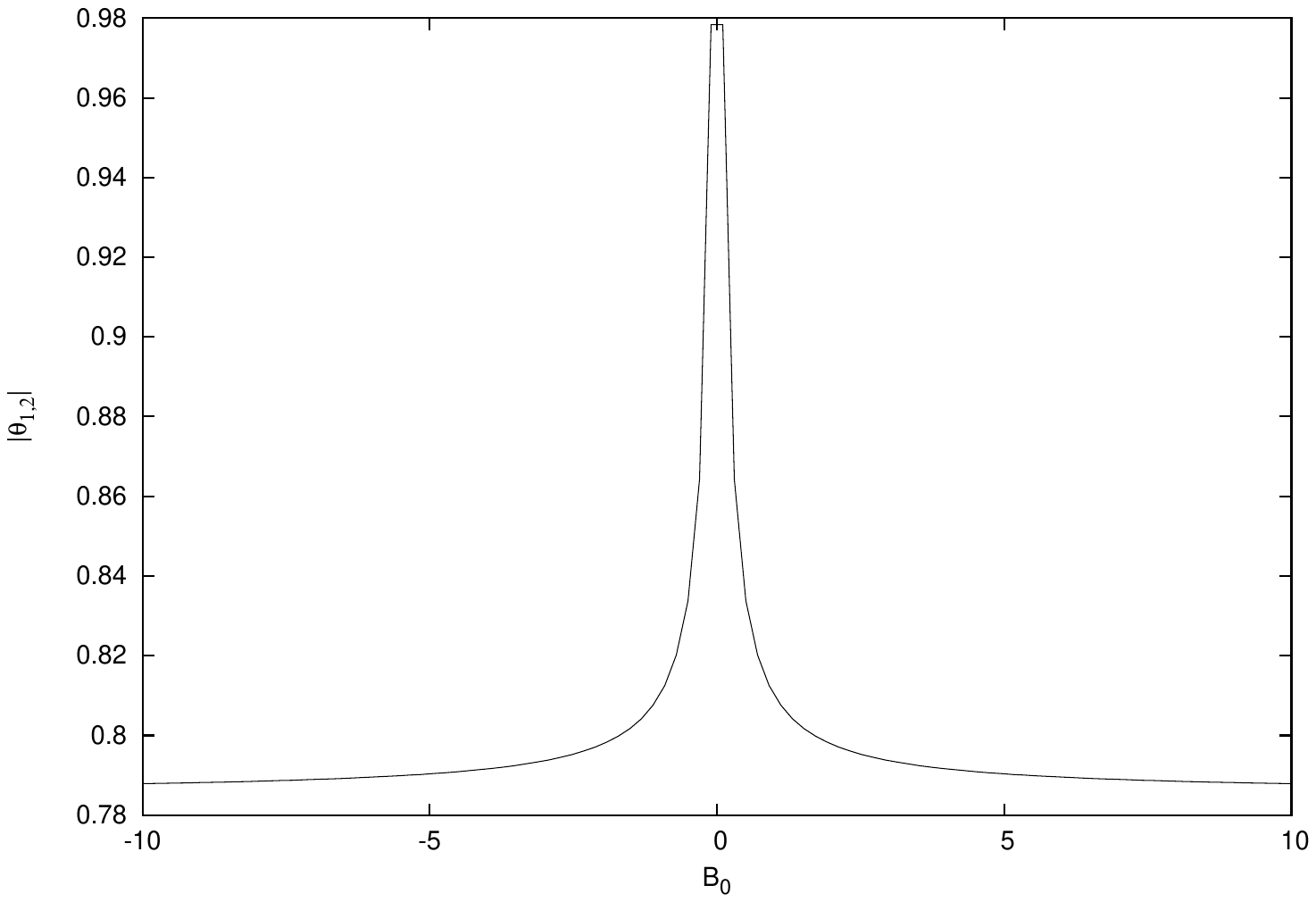}}
        \caption{
Variation of flavor mixing angles in radian as functions of 
gravitational coupling in units of eV with $m_e=0.01$ eV, $m_\mu=0.1$ eV and $m_{\mu e}=0.05$ eV.
}
        \label{figth12}
\end{figure}

\begin{figure}
        \centering
        \fbox{\includegraphics[width=0.78\linewidth]{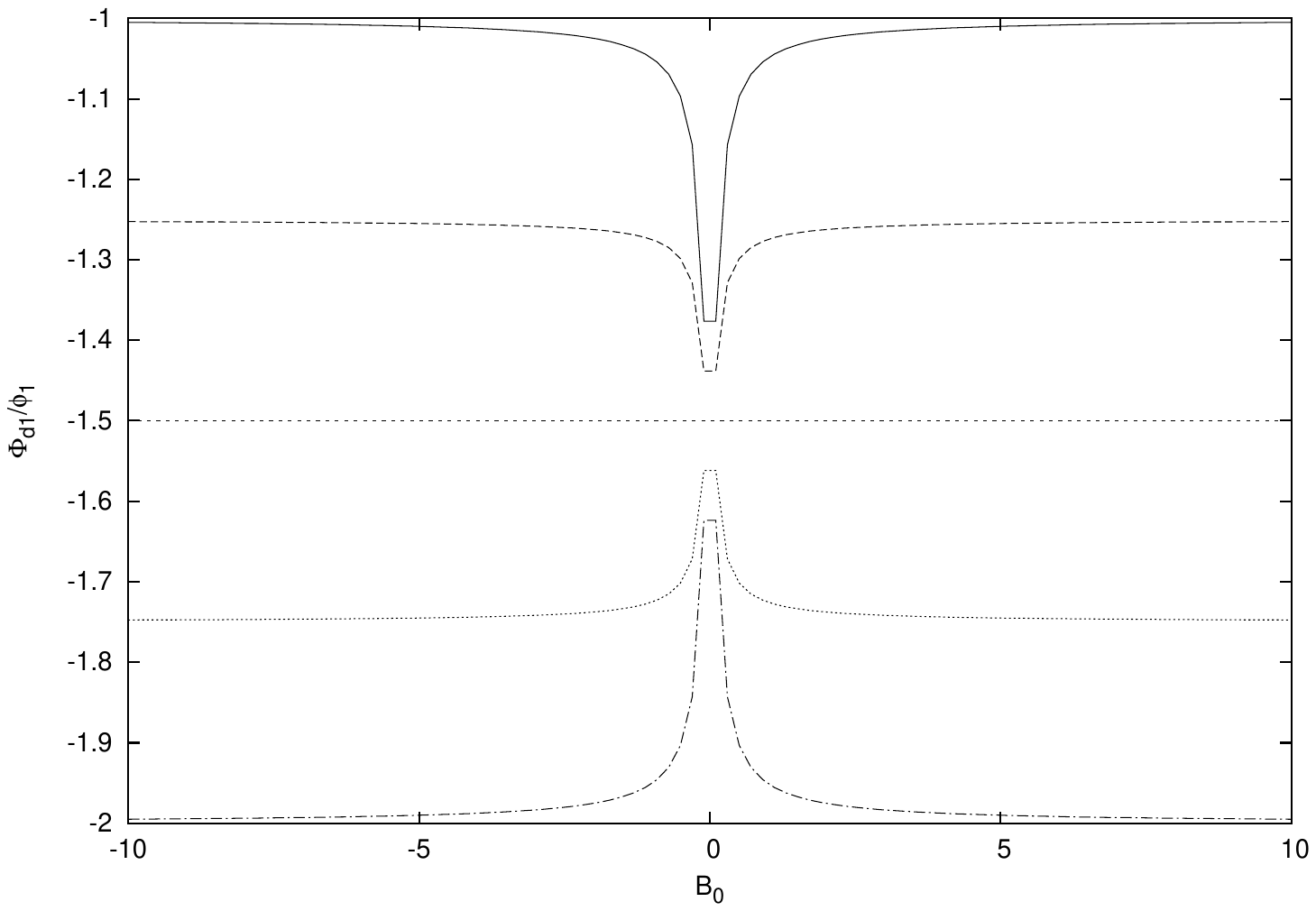}}
        \caption{
Variation of $\tau$-independent part of $\Phi_{d1}$ for entangled states as a function of 
gravitational coupling in units of eV with $m_e=0.01$ eV, $m_\mu=0.1$ eV and $m_{\mu e}=0.05$ eV, 
and $\phi_1=\phi_2=\beta$. From~top to bottom, lines are for $\alpha=0,\pi/6,\pi/4,\pi/3$ and $\pi/2$.
}
        \label{figphen}
\end{figure}

In the presence of very strong gravitational effect ($B_0>>m_e,m_\mu$) (see, e.g.,~\cite{bm1,bm2}),
$\theta_{1,2}\rightarrow\pi/4$ and for entangled states $\Phi_{do1,2}\rightarrow \beta\cos\alpha\sin\alpha$. 
Similarly, the~$\tau$-independent part of $\Phi_{d1,2,3,4}$ 
of entangled states survives even at a very strong gravitational field for an arbitrary $\alpha$.~Figure~\ref{figth12} shows that $\theta_1$ and $\theta_2$ decrease with the increase of $|B_0|$,
which furthermore controls geometric Berry-like
phases associated with entangled states given in Section \ref{sec5}.~Figure~\ref{figphen} shows how
the corresponding $\tau$-independent part of $\Phi_{d1}$ varies with the change of the strength of gravitational coupling. This~confirms that while 
$\theta_{1,2}$ decrease with increasing $B_0$, mixing and also 
various phases still~survive.

\unskip

\unskip

\section{Summary}\label{sec7}

Spinors interacting with background gravity of arbitrary strength in an arbitrary 
spacetime are known to be 
split in energies between up and down spinors and furthermore into the states of 
positive and negative energies. The only requirement is that
the spacetime should not be spherically symmetric.
It~has been shown that such spinors acquire a geometric phase due to background gravitational field 
in the same way as they do 
in a magnetic field. The~necessary condition for such a situation is either the spacetime curvature
coupling to the spinor (gravitational 4-vector potential) is not constant or the momentum of the spinor
is not constant, along with a non-zero (even constant) temporal part of curvature~coupling. 

Neutrinos, as~a class of spinors in nature, are shown to acquire geometric as well as dynamic
phases during their propagation under background gravitational field. 
To have a non-trivial phase induced by the gravitational field of compact 
objects (e.g., black hole), the~
mass of the object producing gravitational fields must not be more than a millionth of a solar mass, {{i.e., primordial in nature. There~are many 
missions to constrain the evidence for primordial black holes, e.g.,~observing specific small interference patterns within gamma-ray bursts by
the \textit{Fermi} Gamma-ray Space Telescope could be the indirect evidence 
for primordial black holes.}}

In the flavor sector, when the background gravity is much stronger than the lepton number violating (Majorana) masses, 
the mixing parameters $\theta_{1,2}$ become a constant equal to $\pi/4$, 
hence~maximum mixing, which could lead to
$\phi$-dependent geometric phases if $\phi$ varies. However, for 
a weak background gravity, $\theta_{1,2}$ are found to depend
on the specific values of the neutrino masses. The~combined neutrino--antineutrino states form an entangled system whose phases 
have also been~calculated. 

\vspace{6pt} 




\acknowledgments{B.M. thanks Subhashish Banerjee of IIT Jodhpur, Kaushik Ghosh of Vivekananda College Kolkata, Tanuman Ghosh of RRI Bangalore, Subroto Mukerjee of IISc Bangalore
for discussions at various stages of the work. 
Thanks are also due to Lars Andersson and Marius Oancea of AEI, Max-Planck 
Institute Potsdam-Golm, for~discussions.  }

\begin{thebibliography}{999}

\bibitem{berry} Berry, M.V. {Quantal phase factors accompanying adiabatic changes.} 
\emph{Proc. R. Soc. A} \textbf{1984}, \emph{392}, 45.

\bibitem{aa} Aharonov, Y.; Anandan, J. {Phase change during a cyclic quantum evolution.} 
\emph{Phys. Rev. Lett.} \textbf{1987}, \emph{58}, 1593. [\href{http://dx.doi.org/10.1103/PhysRevLett.58.1593}{CrossRef}] [\href{http://www.ncbi.nlm.nih.gov/pubmed/10034484}{PubMed}]

\bibitem{sam} {Samuel, J.; Bhandari, R.} 
{General setting for Berry's phase.} 
\emph{Phys. Rev. Lett.} \textbf{1988}, \emph{60}, 2339. [\href{http://dx.doi.org/10.1103/PhysRevLett.60.2339}{CrossRef}] [\href{http://www.ncbi.nlm.nih.gov/pubmed/10038326}{PubMed}]

\bibitem{muksi} Mukunda, N.; Simon, R. {Quantum kinematic approach to the geometric phase. I. General formalism.} 
\mbox{\emph{Ann. Phys. (N. Y.)}} \textbf{1993}, \emph{228}, 205. [\href{http://dx.doi.org/10.1006/aphy.1993.1093}{CrossRef}]

\bibitem{naga} Nakagawa, N. {Geometrical phase factors and higher-order adiabatic approximations.} 
\emph{Ann. Phys. (N. Y.)} \textbf{1987}, \emph{179}, 145. [\href{http://dx.doi.org/10.1016/S0003-4916(87)80007-6}{CrossRef}]

\bibitem{Vidal} Vidal, J.; Wudka, J.  {Non-dynamical contributions to left-right transitions in the solar neutrino problem.} 
\mbox{\emph{Phys. Lett. B}} \textbf{1990}, \emph{249}, 473. [\href{http://dx.doi.org/10.1016/0370-2693(90)91019-8}{CrossRef}]

\bibitem{Aneziris}{Aneziris, C.; Schechter, J. Three Majorana neutrinos in a twisting magnetic field. \emph{Phys. Rev. D} \textbf{1992}, \emph{45}, 1053. [\href{http://dx.doi.org/10.1103/PhysRevD.45.1053}{CrossRef}]

\bibitem{Smirnov}
Smirnov, A.Y. The geometrical phase in neutrino spin precession and the solar neutrino problem. \mbox{\emph{Phys. Lett. B}} \textbf{1991}, \emph{260}, 161. [\href{http://dx.doi.org/10.1016/0370-2693(91)90985-Y}{CrossRef}]

\bibitem{Guzzo}
Guzzo, M.M.; Bellandi, J. On the question of neutrino spin precession in a magnetic field. \emph{Phys. Lett. B} \textbf{1992}, \emph{294}, 243.} 


\bibitem{he} He, X.-G.; Li, X.-Q.; McKellar, B.H.J.; Zhang, Y. {Berry phase in neutrino oscillations.} 
\emph{Phys.
Rev. D} \textbf{2005}, \emph{72},~053012. [\href{http://dx.doi.org/10.1103/PhysRevD.72.053012}{CrossRef}]


\bibitem{blasone} Blasone, M.; Henning, P.A.; Vitiello, G. {Berry phase for oscillating neutrinos.} \emph{Phys. Lett. B} \textbf{1999}, \emph{466}, 262. [\href{http://dx.doi.org/10.1016/S0370-2693(99)01137-5}{CrossRef}]

\bibitem{wang} Wang, X.-B.; Kwek, L.C.; Liu, Y.; Oh, C.H. {Noncyclic phase for neutrino oscillation.} \emph{Phys. Rev.
D} \textbf{2001}, \emph{63},~053003. [\href{http://dx.doi.org/10.1103/PhysRevD.63.053003}{CrossRef}]

\bibitem{sudhir} Joshi, S.; Jain, S. {Geometric phase for neutrino propagation in magnetic field.} \emph{Phys. Lett. B} \textbf{2016}, \emph{754}, 135. [\href{http://dx.doi.org/10.1016/j.physletb.2016.01.023}{CrossRef}]

\bibitem{rajaram} Ramaseshan, S.; Nityananda, R. {The interference of polarized light as an early example of Berry's phase.} \emph{Curr. Sci.} \textbf{1986}, \emph{55}, 1225.

\bibitem{pancha} Pancharatnam, S. {Generalized theory of interference, and its applications.} \emph{Proc. Indian Acad. Sci. A} \textbf{1956}, \emph{44},~247. [\href{http://dx.doi.org/10.1007/BF03046050}{CrossRef}]

\bibitem{mehta} Mehta, P. {Topological phase in two flavor neutrino oscillations.} \emph{Phys. Rev. D} \textbf{2009}, \emph{79}, 096013. [\href{http://dx.doi.org/10.1103/PhysRevD.79.096013}{CrossRef}]

\bibitem{new1} Dajka, J.; Syska, J.; Łuczka, J. {Geometric phase of neutrino propagating through dissipative matter.} \emph{\mbox{Phys. Rev. D}} \textbf{2011}, \emph{83}, 097302. [\href{http://dx.doi.org/10.1103/PhysRevD.83.097302}{CrossRef}]

\bibitem{new2} Syska, J.; Dajka, J.; Łuczka, J. {Interference phenomenon and geometric phase for Dirac neutrino in $\pi^+$ decay.} \emph{Phys. Rev. D} \textbf{2013}, \emph{87}, 117302. [\href{http://dx.doi.org/10.1103/PhysRevD.87.117302}{CrossRef}]

\bibitem{new3} Joshi, S.; Jain, S.R. {Noncyclic geometric phases and helicity transitions for neutrino oscillations in a magnetic field.} \emph{Phys. Rev. D} \textbf{2017}, \emph{96}, 096004. [\href{http://dx.doi.org/10.1103/PhysRevD.96.096004}{CrossRef}]

\bibitem{new4} Johns, L.; Fuller, G.M. {Geometric phases in neutrino oscillations with nonlinear refraction.} \emph{Phys. Rev. D} \textbf{2017}, \emph{95}, 043003. [\href{http://dx.doi.org/10.1103/PhysRevD.95.043003}{CrossRef}]

\bibitem{new5} Wang, Z.; Pan, H. {Exploration of CPT violation via time-dependent geometric quantities embedded in neutrino oscillation through fluctuating matter.} \emph{Nuc. Phys. B} \textbf{2017}, \emph{915}, 414. [\href{http://dx.doi.org/10.1016/j.nuclphysb.2016.12.019}{CrossRef}]

\bibitem{new6} Dixit, K.; Alok, A.K.; Banerjee, S.; Kumar, D. {Geometric phase and neutrino mass hierarchy problem.}  \mbox{\emph{J. Phys. G}} \textbf{2018}, \emph{45}, 085002. [\href{http://dx.doi.org/10.1088/1361-6471/aac454}{CrossRef}]

\bibitem{new7} Capolupo, A.; Giampaolo, S.M.; Hiesmayr, B.C.; Vitiello, G. {Geometric phase of neutrinos: Differences between Dirac and Majorana neutrinos.}
\emph{Phys. Lett. B} \textbf{2018}, \emph{780}, 216. [\href{http://dx.doi.org/10.1016/j.physletb.2018.03.016}{CrossRef}]

\bibitem{new8} Simonov, K.; Capolupo, A.; Giampaolo, S.M. {Gravity, entanglement and CPT-symmetry violation in particle mixing.} \emph{Eur. Phys.
J. C} \textbf{2019}, \emph{79}, 902. [\href{http://dx.doi.org/10.1140/epjc/s10052-019-7407-y}{CrossRef}]

\bibitem{bm1} Mukhopadhyay, B. {Gravity-induced neutrino-antineutrino oscillation:~CPT and lepton number non-conservation under gravity.} \emph{Class. Quantum Gravity} \textbf{2007}, \emph{24}, 1433. [\href{http://dx.doi.org/10.1088/0264-9381/24/6/004}{CrossRef}]

\bibitem{bm2} Sinha, M.; Mukhopadhyay, B. {CPT and lepton number violation in the neutrino sector: Modified mass matrix and oscillation due to gravity.} \emph{Phys. Rev. D} \textbf{2008}, \emph{77}, 025003. [\href{http://dx.doi.org/10.1103/PhysRevD.77.025003}{CrossRef}]

\bibitem{param} Singh, P.; Mukhopadhyay, B. {Gravitationally induced neutrino asymmetry.} \emph{Mod. Phys. Lett. A} \textbf{2003}, \emph{18}, 779. [\href{http://dx.doi.org/10.1142/S0217732303009691}{CrossRef}]

\bibitem{bmmpla} Mukhopadhyay, B. {Neutrino asymmetry around black holes: Neutrinos interact with gravity.} \emph{Mod. Phys. Lett.~A} \textbf{2005}, \emph{20}, 2145. [\href{http://dx.doi.org/10.1142/S0217732305017640}{CrossRef}]

\bibitem{ujjal} Debnath, U.; Mukhopadhyay, B.; Dadhich, N. { Spacetime Curvature Coupling of Spinors in Early Universe:. Neutrino Asymmetry and a Possible Source of Baryogenesis.} \emph{Mod.
Phys. Lett. A} \textbf{2006}, \emph{21}, 399. [\href{http://dx.doi.org/10.1142/S0217732306019542}{CrossRef}]


\bibitem{kim} Kim, C.W.; Sze, W.K.; Nussinov, S. {Neutrino oscillations and the Landau-Zener formula.} \emph{Phys. Rev. D} \textbf{1987}, \emph{35},
4014. [\href{http://dx.doi.org/10.1103/PhysRevD.35.4014}{CrossRef}]

\bibitem{stodo} Stodolsky, L. {Treatment of neutrino oscillations in a thermal environment.} \emph{Phys. Rev. D} \textbf{1987}, \emph{36}, 2273. [\href{http://dx.doi.org/10.1103/PhysRevD.36.2273}{CrossRef}] [\href{http://www.ncbi.nlm.nih.gov/pubmed/9958431}{PubMed}]

\bibitem{wudka} P\'iriz, D.;  Roy, M.; Wudka, J. {Neutrino oscillations in strong gravitational fields.} \emph{Phys. Rev. D} \textbf{1996}, \emph{54}, 1587. [\href{http://dx.doi.org/10.1103/PhysRevD.54.1587}{CrossRef}] [\href{http://www.ncbi.nlm.nih.gov/pubmed/10020833}{PubMed}]

\bibitem{bm02}Mohanty, S.; Mukhopadhyay, B.; Prasanna, A.R. {Experimental tests of curvature couplings of fermions in general relativity.} \emph{Phys. Rev. D} \textbf{2002}, \emph{65}, 122001. [\href{http://dx.doi.org/10.1103/PhysRevD.65.122001}{CrossRef}]

\bibitem{khushbooBM} Dixit, K.; Naikoo, J.; Mukhopadhyay, B.; Banerjee, S. {Quantum correlations in neutrino oscillations in curved spacetime.}
\emph{Phys. Rev. D} \textbf{2019}, \emph{100}, 055021. [\href{http://dx.doi.org/10.1103/PhysRevD.100.055021}{CrossRef}]


\bibitem{birrel} Birrell, N.D.; Davies, P. \emph{Quantum Fields in Curved Space}; {Cambridge University Press:} 
Cambridge, UK, 1982.

\bibitem{kaku} Kaku, M. \emph{Quantum Field Theory}; Oxford University Press: Oxford, UK, 1993.



\bibitem{schw} Schwinger, J. \emph{Particles, Sources, and Fields III}; Addison-Wesley: {Redwood City, CA, USA,} 
1989.

\bibitem{obukhov} Obukhov, Y.N. {Spin, Gravity, and Inertia.} \emph{Phys. Rev. Lett.} \textbf{2001}, \emph{86}, 192. [\href{http://dx.doi.org/10.1103/PhysRevLett.86.192}{CrossRef}]

\bibitem{parker09} {Huang, X.; Parker, L. {Hermiticity of the Dirac Hamiltonian in curved spacetime.} \emph{Phys. Rev. D} \textbf{2009}, \emph{79}, 024020.} [\href{http://dx.doi.org/10.1103/PhysRevD.79.024020}{CrossRef}]

\bibitem{bookmy} Mukhopadhyay, B. Exploring the Universe: From Near Space to Extra-Galactic.
In \emph{Astrophysics and Space Science Proceedings}; Mukhopadhyay, B., Sasmal, S., Eds.; Springer: {Berlin/Heidelberg, Germany,} 
2018; Volume 53, p. 3.

\bibitem{kost} Colladay, D.; Kosteleck\'y, V.A. {CPT violation and the standard model.} \emph{Phys. Rev. D} \textbf{1997}, \emph{55}, 6760. [\href{http://dx.doi.org/10.1103/PhysRevD.55.6760}{CrossRef}]

\bibitem{nick1} Ellis, J.; Mavromatos, N.E. {Role of space-time foam in breaking supersymmetry via the Barbero-Immirzi parameter.} \emph{Phys. Rev. D} \textbf{2011}, \emph{84}, 085016. [\href{http://dx.doi.org/10.1103/PhysRevD.84.085016}{CrossRef}]

\bibitem{nick2} Mavromatos, N.E.; Sarkar, S. {CPT-violating leptogenesis induced by gravitational defects.} \emph{Eur. Phys. J. C} \textbf{2013}, \emph{73}, 2359. [\href{http://dx.doi.org/10.1140/epjc/s10052-013-2359-0}{CrossRef}]

\bibitem{harm} Mosquera Cuesta, H.J. {Neutrino astrophysics in slowly rotating spacetimes permeated by nonlinear electrodynamics fields.}  \emph{Astrophys. J.} \textbf{2017}, \emph{835}, 215. [\href{http://dx.doi.org/10.3847/1538-4357/aa5266}{CrossRef}]


\bibitem{exp} Diaz, J.S.;  Katori, T.; Spitz, J.; Conrad, J.M. {
Search for neutrino-antineutrino oscillations with a reactor experiment.} \emph{Phys. Lett. B} \textbf{2013}, \emph{727}, 412. [\href{http://dx.doi.org/10.1016/j.physletb.2013.10.058}{CrossRef}]

\bibitem{st} Shapiro, S.L.; Teukolsky, S.A. {\it Black Holes, White Dwarfs, and Neutron
Stars: The Physics of Compact Objects}; John~Wiley \& Sons: New York, NY, USA, 1983.

\bibitem{eriks} Sj\"oqvist, E.  { Geometric phase for entangled spin pairs.} \emph{Phys. Rev. A} \textbf{2000}, \emph{62}, 022109. [\href{http://dx.doi.org/10.1103/PhysRevA.62.022109}{CrossRef}]

\end{thebibliography}
\end{document}